\documentclass[prl,twocolumn,a4paper,superscriptaddress,showpacs]{revtex4}

\usepackage{amsmath}
\usepackage{graphicx}
\usepackage{sistyle}
\usepackage{todonotes}
\usepackage{gensymb}
\usepackage{bbold}
\usepackage{bm}
\usepackage{natbib}
\usepackage{amsfonts}
\usepackage{dsfont}
\usepackage{amssymb}
\usepackage{latexsym}
\usepackage{fancyhdr}
\usepackage[bbgreekl]{mathbbol}
\usepackage{epsfig}
\usepackage{color}
\usepackage{textcomp}

\def\bra#1{\mathinner{\langle{#1}|}}
\def\ket#1{\mathinner{|{#1}\rangle}}

\def\text#1{\textrm{#1}}

\DeclareMathOperator{\Tr}{Tr}

\begin{document}

\title{Optomechanical Bell test}

\date{\today}
\author{V.~Caprara Vivoli}
\affiliation{Group of Applied Physics, University of Geneva, CH-1211 Geneva 4, Switzerland}
\author{T.~Barnea}
\affiliation{Group of Applied Physics, University of Geneva, CH-1211 Geneva 4, Switzerland}
\author{C.~Galland}
\affiliation{Ecole Polytechnique Federale de Lausanne (EPFL), CH-1015 Lausanne, Switzerland}
\author{N.~Sangouard}
\affiliation{Department of Physics, University of Basel, CH-4056 Basel, Switzerland}

\date{\today}

\begin{abstract}
Photons of a laser beam driving the upper motional sideband of an optomechanical cavity can decay into photon-phonon pairs by means of an optomechanical parametric process. The phononic state can subsequently be mapped to a photonic state by exciting the lower sideband, hence creating photon-photon pairs out of an optomechanical system. We here demonstrate that these pairs can violate a Bell inequality when they are measured with photon counting techniques preceded by small displacement operations in the phase space. Our results show how to detect non-local light-mechanical correlations and open the way for device-independent quantum optomechanics where non-classical light-mechanical correlations can be certified without assumptions on the state structure or on the measurement devices.
\end{abstract}
\maketitle

\paragraph{Introduction ---}
Cavity optomechanics which describes a mechanical oscillator controlled by an electromagnetic cavity mode via a generalized radiation pressure force, is the subject of intense research \cite{Kippenberg08, Favero09, Marquardt09, Meystre13, Aspelmeyer14}. Most recent progress includes the cooling of mechanical oscillators \cite{Gigan06, Arcizet06, Kleckner06, Schliesser06, Schliesser09, Vanner13} down to the ground state \cite{O'Connell10, Teufel11, Chan11, Safavi-Naeini12}, the readout of the mechanical position with a readout imprecision below the standard quantum limit \cite{Teufel09, Anetsberger10} as well as optomechanical squeezing \cite{Szorkovszky13} and entanglement \cite{Palomaki13}. Reciprocally, the mechanical degrees of freedom can be used to control the cavity light e.g. for fast and slow light \cite{Safavi-Naeini11, Zhou13}, frequency conversions \cite{Hill12, Bochmann13, Andrews14}, squeezing \cite{Safavi-Naeini13} and information storage in long-lived mechanical oscillations \cite{Palomaki13, Fiore11}. 

Optomechanical systems are also envisioned as test-benches for physical theories \cite{Bose97, Bose99, Marshall03, Kleckner08, Pepper12, Sekatski14, Ghobadi14, Bahrami14, Nimmrichter14, Diosi15, Ho15}. As a step in this direction, quantum correlations between light and mechanics have been observed recently \cite{Palomaki13}. In this experiment, quantum features have been detected through an entanglement witness where one assumes that the measurement devices are well characterized and where quantum theory is used  to predict the results of these measurements on separable states. It is interesting to wonder whether the non-classical behavior of optomechanical systems can be certified outside of the quantum formalism, i.e. from a Bell test \cite{Brunner14}. This is particularly relevant to test post-quantum theories including explicit collapse models \cite{Ellis84, GRW86, GRW90, Gisin89, Diosi89, Penrose96}, where the assumption that the system behaves quantum-mechanically may be questionable \cite{Pfister15}. 

In this Letter, we show how to perform such a Bell test in the experimentally relevant weak-optomechanical coupling and sideband-resolved regime. Our proposal, which starts with a mechanical oscillator close to its ground state, consists of two steps. First, the optomechanical system is excited by a laser tuned to the upper motional sideband of the cavity to create photon-phonon pairs via optomechanical parametric conversion. Second, a laser resonant with the lower sideband is used to map the phononic state to the cavity field. The correlations between the photons generated at the cavity frequency during the first and second steps are then analyzed by photon counting preceded by small displacement operations in the phase space. We show that they violate the Bell-CHSH inequality \cite{CHSH69}, revealing two strong properties of the optomechanical system. First, under the assumption that one of the photonic mode is a faithful representative of the phononic mode, we conclude \textit{outside the quantum formalism} that the \textit{photon-phonon correlations are non-local}, i.e. they cannot be reproduced by local-hidden variables \cite{Brunner14}. Second, such a violation certifies \textit{within the quantum formalism that the optomechanical state is non-local, i.e. provides stronger correlations than entanglement} \cite{Werner89}. This claim is \textit{device-independent} i.e. holds without assumptions on the dimension of the underlying Hilbert space or on the functioning of the measurement devices \cite{Scarani12}.  
Very recent experiments \cite{Cohen15} performing photon counting in optomechanical experiments promote the feasibility of our approach. \\

\paragraph{Principle of the optomechanical Bell test ---} 
The basic principle is inspired by Ref. \cite{Galland14}. We use two laser pulses driving either the upper or the lower optomechanical sideband, at frequency $\omega_\pm$ which is the sum or the difference of the cavity ($\omega_c$) and the mechanical ($\Omega_m$) frequencies. The optomechanical Hamiltonian includes $H_0=\hbar \omega_c a^{\dagger}a+\hbar \Omega_m b^{\dagger}b$ the uncoupled cavity and mechanical systems with respective bosonic operators $a$ and $b,$ $H_{\text{OM}}=-\hbar g_0a^{\dagger}a(b^{\dagger}+b)$ the optomechanical interaction with $g_0$ the optomechanical coupling, and $H_l=\hbar(s^{*}_\pm e^{i\omega_{\pm}t}a+s_\pm e^{-i\omega_{\pm}t}a^{\dagger})$ the driving laser with $|s_\pm|=\sqrt{\kappa P_\pm /\hbar \omega_\pm},$ $P_\pm$ being the laser power and $\kappa$ the cavity decay rate (assuming that the intracavity loss is negligible). In the interaction picture, the weak coupling limit $g_0 \ll \kappa$ and the resolved sideband regime $\kappa \ll \Omega_m,$ the dynamics are given by a set of Langevin equations 
\begin{eqnarray}
\label{cavity}
&&\frac{d a}{dt} = \frac{i}{\hbar} [H_\pm,a] - \frac{\kappa}{2} a + \sqrt{\kappa} a_{\text{in}} \\
\label{meca}
&&\frac{d b}{dt} = \frac{i}{\hbar} [H_\pm,b] - \frac{\gamma}{2} b + \sqrt{\gamma} b_{\text{in}}
\end{eqnarray}
with the linearized Hamiltonians $H_+= - g_+ \hbar a^\dag b^\dag + h.c.$ for a blue detuned drive and $H_-= - g_-\hbar a^\dag b + h.c.$ for a red detuned drive. $g_\pm$ are the effective optomechanical coupling rates enhanced by the intra-cavity photon number $g_\pm= g_0 \sqrt{n_\pm} = \frac{\kappa P_\pm}{\hbar \omega_c (\Omega_m^2+ \kappa^2/4)}.$ $a_{in}$ is the vacuum noise entering the cavity. We assume that the laser is shot-noise limited, hence does not add contributions to the input noise. $b_{in}$ is the thermal noise from a phonon bath at temperature $T_{\text{bath}}$ and mean occupation number $n_{\text{th}}=\frac{k_B T_{\text{bath}}}{\hbar \Omega_m}.$ In the following treatment, we neglect the mechanical decay which is well justified for timescales smaller than the thermal decoherence time $1/\gamma n_{\text{th}},$ $\gamma$ being the coupling rate between the mechanical oscillator and the thermal bath. \\

Consider first the case where the mechanics is driven by a blue detuned laser. In the regime $g_+ \ll \kappa,$ the cavity mode can be adiabatically eliminated and Eq. \eqref{cavity} leads to $a_1 = \frac{2}{\kappa} (i g_+ b^\dag + \sqrt{\kappa} a_{1,\text{in}})$ (the subscript on the cavity field operators is used to recall that we are considering the first step). Further introducing the input/output relation \cite{Gardiner85}, $a_{1,\text{out}} = - a_{1,\text{in}} + \sqrt{\kappa} a_1,$ we obtain 
\begin{eqnarray}
\label{cavity_1}
&& a_{1,\text{out}} = a_{1,\text{in}} + i \sqrt{2 \bar{g}_+} b^\dag,\\
\label{meca_1}
&& \frac{d b}{dt} = \bar{g}_+ b + i \sqrt{2 \bar{g}_+} a^{\dag}_{1,\text{in}}
\end{eqnarray}
where $\bar{g}_+ = \frac{2 g_+^2}{\kappa}.$ We then follow Hofer et al. \cite{Hofer11} and introduce the temporal modes 
$
A_{1,\text{in}}(t)=\sqrt{\frac{2 \bar{g}_+}{1-e^{-2\bar{g}_+ t}}}\int_0^{t} dt' e^{-\bar{g}_+ t'}a_{1,\text{in}}(t'),
$
$
A_{1,\text{out}}(t)=\sqrt{\frac{2 \bar{g}_+}{e^{2\bar{g}_+ t}-1}}\int_0^{t} dt' e^{\bar{g}_+ t'}a_{1,\text{out}}(t').
$
The solutions of Eqs. \eqref{cavity_1} -- \eqref{meca_1} take the following simple forms
$A_{1,\text{out}}(t)=e^{\bar{g}_+ t}A_{1,\text{in}}(t)+i\sqrt{e^{2\bar{g}_+ t}-1}b^{\dagger}(0),$
$b(t)=e^{\bar{g}_+ t}b(0)+i\sqrt{e^{2\bar{g}_+ t}-1}A_{1,\text{in}}^{\dagger}(t).$
These solutions can be rewritten as  $A_{1,\text{out}}=\tilde{U}_1^{\dag}(t)A_{1,\text{in}}\tilde{U}_1(t)$ and $b(t)=\tilde{U}_1^{\dagger}(t)b(0)\tilde{U}_1(t)$ where the propagator is given by
\begin{eqnarray}
\label{Utilde}
&\tilde{U}_1(t)&=e^{i\sqrt{1-e^{-2 \bar{g}_+ t}}A_{1,\text{in}}^{\dagger} b^{\dagger}}\\
\nonumber
&&\times e^{\bar{g}_+ t(-1-A_{1,\text{in}}^{\dagger}A_{1,\text{in}}-b^{\dagger}b)}e^{i\sqrt{1-e^{-2 \bar{g}_+ t}}A_{1,\text{in}}b}.
\end{eqnarray}
When applied on the vacuum, this propagator leads to the creation of photon-phonon pairs where the number of photons equals the number of phonons, each of them following a thermal statistics with mean excitation number ${e^{2 \bar{g}_+ t}-1}$.\\

Now consider the case where the mechanics is driven by a red detuned laser, i.e. the dynamics is given by the beam-splitter Hamiltonian $H_-$. Following the same procedure as before, Eqs. \eqref{cavity_1} -- \eqref{meca_1} become
\begin{eqnarray}
\label{cavity_2}
&& a_{2,\text{out}} = a_{2,\text{in}} + i \sqrt{2 \bar{g}_-} b,\\
\label{meca_2}
&& \frac{d b}{dt} = -\bar{g}_- b + i \sqrt{2 \bar{g}_-} a_{2,\text{in}}
\end{eqnarray}
 where $\bar{g}_- = \frac{2 g_-^2}{\kappa}.$ Introducing the modes 
$
A_{2,\text{in}}(t)=\sqrt{\frac{2 \bar{g}_-}{e^{2\bar{g}_- t}-1}}\int_0^{t} dt' e^{\bar{g}_- t'}a_{2,\text{in}}(t'),
$
$
A_{2,\text{out}}(t)=\sqrt{\frac{2 \bar{g}_-}{1-e^{-2\bar{g}_- t}}}\int_0^{t} dt' e^{-\bar{g}_- t'}a_{2,\text{out}}(t')
$
leads to the simple expression for the solutions of Eqs. \eqref{cavity_2}-\eqref{meca_2} at a time $t$ after the beginning of the red detuned pulse $A_{2,\text{out}}(t)=e^{-\bar{g}_- t}A_{2,\text{in}}(t)+i\sqrt{1-e^{-2\bar{g}_- t}}b(0),$ $b(t)=e^{-\bar{g}_- t}b(0)+i\sqrt{1-e^{-2\bar{g}_- t}}A_{2,\text{in}}(t).$
These solutions can be rewritten as  $A_{2,\text{out}}=\tilde{U}_2^{\dag}(t)A_{2,\text{in}}\tilde{U}_2(t)$ and $b(t)=\tilde{U}_2^{\dagger}(t)b(0)\tilde{U}_2(t)$ where the propagator is given by
\begin{eqnarray}
\label{Utilde}
&\tilde{U}_2(t)&=e^{i\sqrt{e^{2\bar{g}_- t }-1}A_{2,\text{in}}b^{\dagger}}  \\
\nonumber
&& \times e^{-\bar{g}_- t(A_{2,\text{in}}^\dag A_{2,\text{in}} - b^\dag b)} e^{i\sqrt{e^{2\bar{g}_- t }-1}A_{2,\text{in}}^{\dagger} b}.
\end{eqnarray}
This corresponds to a process converting a phonon into a photon with probability $1-e^{-2 \bar{g}_- t}$. \\

Now consider an initial state where both optical modes $A_1$ and $A_2$ are empty and where the mechanics is prepared in its ground state. Switching on the blue detuned laser for a time $T_1,$ then the red detuned laser for a time $T_2$ leads to a photon-photon state in mode $A_{1,\text{out}},A_{2,\text{out}}$ given by $\rho_{A_1,A_2}={\Tr}_b \tilde{U}^2(T_2)\tilde{U}^1(T_1) |0,0,0\rangle_{A_{1,\text{in}},A_{2,\text{in}},b} \langle 0,0,0| \tilde{U}^{1 \dag}(T_1)\tilde{U}^{2 \dag}(T_2).$ In the ideal limit $\bar g_- T_2 \rightarrow +\infty,$ the phonon-photon mapping is perfect and the state $\rho_{A_1,A_2}$ corresponds to a two-mode squeezed vacuum. In the general case where $\bar g_- T_2$ has a finite value, $\rho_{A_1,A_2}$ still corresponds to a squeezed vacuum but where  the mode $A_{2}$ undergoes loss. This loss can be modeled by a beam-splitter with a transmission $T=1-e^{-2 \bar{g}_- t}.$ The next section shows how to reveal the non-local content of such a state.\\

In order to test a Bell inequality with the modes $A_{1,\text{out}},A_{2,\text{out}},$ (the subscript $\texttt{\textquotesingle \text{out}\textquotesingle}$ is omitted below) we consider a single-photon detector -- which does not resolve the photon number -- combined with a displacement operation $D(\alpha).$ We associate the outcomes $+1$ / $-1$ to the absence of detection / to the detection of at least one photon. In the subspace composed of the vacuum and the single photon Fock state, such a measurement corresponds exactly to the observable $\sigma_z$ for $\alpha=0$ while for $\alpha=1$ ($\alpha=i$), it is a noisy $\sigma_x$ ($\sigma_y$) \cite{Caprara15b}. The potential of such measurements for non-locality detection has been highlighted in Refs. \cite{Banaszek98, Hessmo04, Chavez11, BohrBrask13, Seshadreesan13}. Refs. \cite{Kuzmich00, Lee09, Bohr12} have also shown how they can be used for Bell tests in photonic experiments where two-mode squeezed states are produced through spontaneous parametric down conversion. More recently, they have been used to reveal genuine path entanglement \cite{Monteiro15}. Further note that a displacement is easy to implement in practice as it requires a coherent state and an unbalanced beamsplitter only \cite{Paris96}.\\

The joint probability $P(+1+1| \alpha_{1} \alpha_{2})$ to get the outcomes $+1$ for both $A_1$ and $A_2$ when they are analyzed with photon counting with efficiency $\eta$ preceded by displacements with amplitude $\alpha_1$ and $\alpha_2$ for $A_1$ and $A_2$ respectively is given by 
$
P(+1+1| \alpha_{1} \alpha_{2}) = \Tr \left(\rho_{A_1,A_2} \mathcal{O}_{\eta}(\alpha_1, A_1) \otimes \mathcal{O}_{\eta}(\alpha_2, A_2)\right)
$
where $\mathcal{O}_{\eta}(\alpha_i, A_i) = D^{\dagger}(\alpha_i)(1-\eta)^{A_i^{\dagger}A_i}D(\alpha_i).$ Such a probability can be computed easily by noting that loss and displacement can be commuted by changing the amplitude of the displacement. In particular 
$
P(+1+1| \alpha_{1} \alpha_{2}) = \Tr \left(\bar \rho_{A_1,A_2} \mathcal{O}_{\eta}(\alpha_1, A_1) \otimes \mathcal{O}_{\eta'}(\alpha_2 \sqrt{T}, A_2)\right)
$
where $\eta'=\eta T,$ and $\bar \rho_{A_1,A_2} = (1-p) e^{-\sqrt{p}A_1^{\dagger}A_2^{\dagger}}\ket{0}_{A_1, A_2} \bra{0} e^{-\sqrt{p}A_1 A_2}$ is simply the two mode squeezed vacuum. ($1-p=e^{-2 \bar g_+ T_1}$ is the probability that both modes are empty.) We find
\begin{figure}
\includegraphics[width=0.4\textwidth]{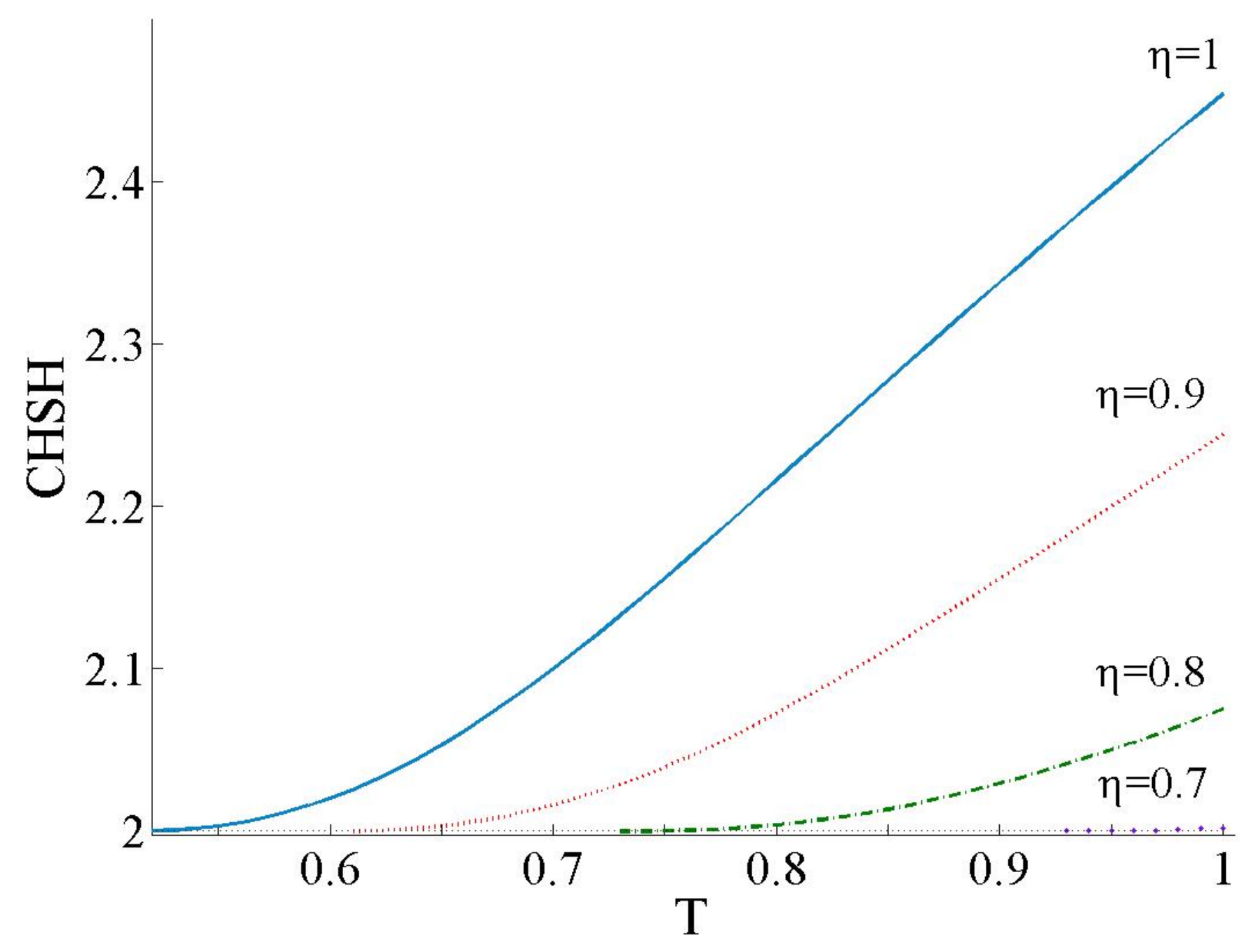}
\caption{CHSH values optimized over the measurement settings $(\alpha_1, \alpha_2)$ as a function of the optomechanical mapping efficiency $(T=1-e^{-2 \bar{g}_- t})$ for various detection efficiencies $\eta.$ The CHSH value is larger than the local bound 2 for unit detection efficiencies when $T \geq 52$\% while for unit phonon-photon mapping efficiency, $\eta \geq 66.8$\% is required.} 
\label{Fig1}
\end{figure} 
\begin{eqnarray}
\label{joint_proba}
&&P(+1+1| \alpha_{1} \alpha_{2})=\frac{(1-p)}{1-p(1-\eta) (1-\eta')} \times \\
\nonumber
&& e^{-\frac{\eta |\alpha_1|^2 (1-(1-\eta') p)+\eta' |\alpha_2|^2 T (1-(1-\eta) p)}{1-p(1-\eta) (1-\eta')}} 
e^{\frac{\eta \eta' \sqrt{p} (\alpha_1^* \alpha_2^* +\alpha_1 \alpha_2)\sqrt{T}}{1-p(1-\eta) (1-\eta')}}.
\end{eqnarray}
Together with the marginals 
\begin{eqnarray}
\nonumber
&&P(+1|\alpha_1)=\frac{(1-p)}{1-p(1-\eta)} \times e^{-\frac{\eta(1-p)|\alpha_1|^2}{1-p(1-\eta)}}\\
\nonumber
&&P(+1|\alpha_2)=\frac{(1-p)}{1-p(1-\eta')} \times e^{-\frac{\eta'(1-p)|\alpha_2|^2T}{1-p(1-\eta')}}
\end{eqnarray}
we get the explicit value of the correlator 
$
E^{\alpha_1,\alpha_2} = 1-2 (P(+1|\alpha_1) + P(+1|\alpha_2)) + 4 P(+1+1| \alpha_{1} \alpha_{2})
$
to test the Bell-CHSH inequality 
$
\text{CHSH}=|E^{\alpha_1,\alpha_2}+E^{\alpha'_1,\alpha_2}+E^{\alpha_1,\alpha'_2}-E^{\alpha'_1,\alpha'_2}| \leq 2
$
which holds for any local hidden-variable model. 

Fig. \ref{Fig1} shows the CHSH values obtained from the optimization over the measurement settings $\alpha_i, \alpha'_i,$ $i \in [1,2]$ as a function of the photon-phonon mapping efficiency $T=1-e^{-2 \bar{g}_- t}$ for various detection efficiencies $\eta$. For high enough efficiencies, we see that the CHSH inequality is violated, hence showing that the correlations of modes $A_1$ and $A_2$ cannot be reproduced by local hidden-variable theories. 

\begin{figure}
\includegraphics[width=0.4\textwidth]{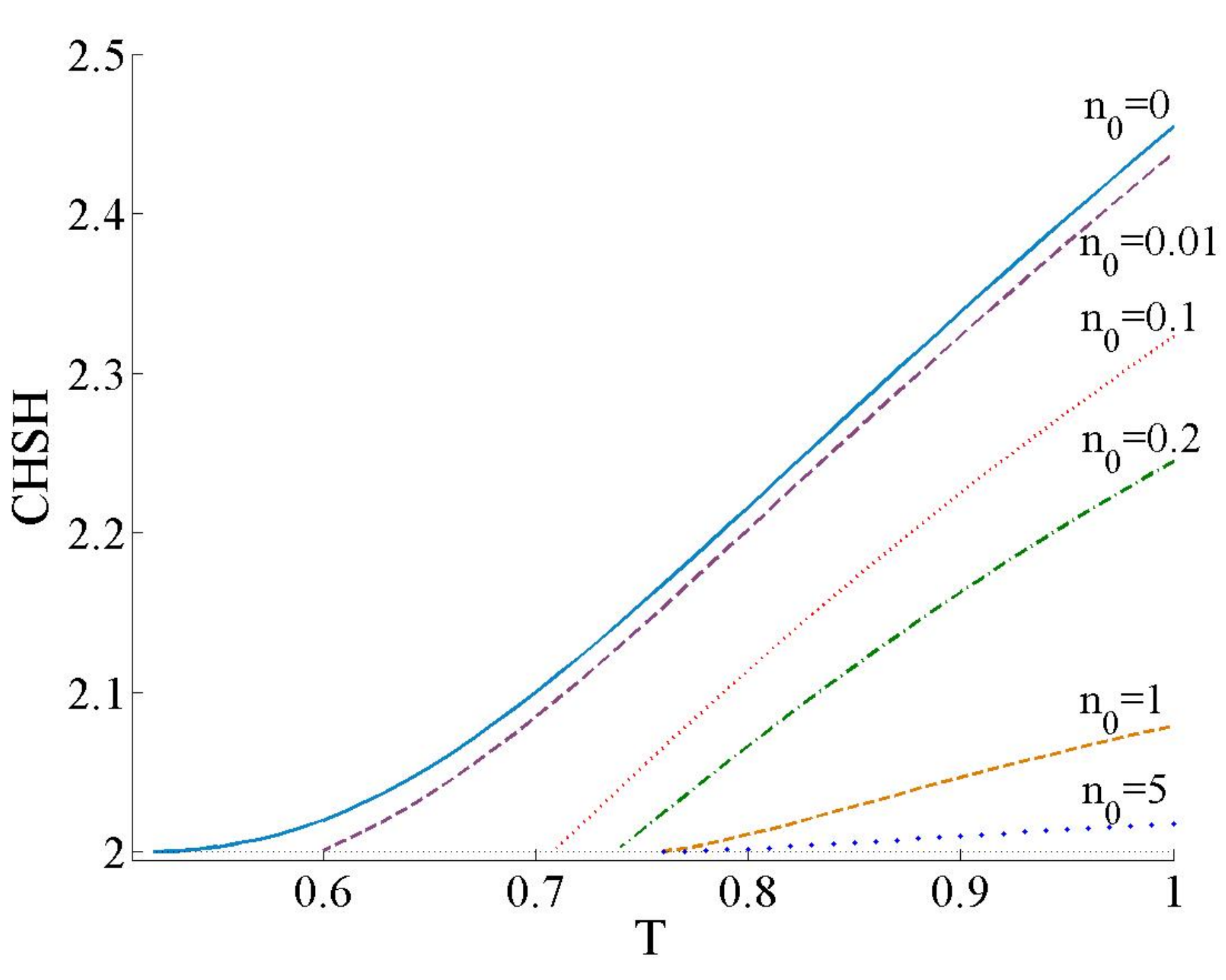}
\caption{CHSH values as a function of the optomechanical mapping efficiency $(T=1-e^{-2 \bar{g}_- t})$ for various mechanical occupation numbers $(n_0)$ assuming unit detection efficiencies $(\eta=1).$} 
\label{Fig2}
\end{figure}

In the above discussion, we have assumed that the mechanical system is prepared in its ground state. In the more general case where the mechanical system is prepared in a thermal state with mean occupation number $n_0,$ the expressions of the joint probability $P(+1+1| \alpha_{1} \alpha_{2})$ and the marginals $P(+1|\alpha_i)$ can be derived as before, c.f. Appendix. The CHSH values resulting from the optimization over the measurement settings are given in Fig. \ref{Fig2} as a function of the phonon-photon mapping efficiency for various mean mechanical occupation numbers assuming unit detection efficiencies. A substantial violation can be obtained if $n_0 \ll 1.$\\

\paragraph{Feasibility ---}
In this paragraph, we discuss the experimental feasibility of the proposed Bell test in detail. The requirements for detecting non-locality are (i) sideband regime $\kappa \ll \Omega_m,$ (ii) weak coupling regime $g_0 \ll \kappa,$ (iii)  ground state cooling $n_0 \ll 1$ and negligible mechanical decoherence during the durations of blue and red detuned laser pulses $T_1+T_2 \ll (\gamma n_{\text{th}})^{-1}.$ Given that $n_{\text{th}} \propto \Omega_m^{-1},$ these conditions are easier to meet with high Q and high frequency $\Omega_m$ oscillators. 
While superconducting microwave optomechanical cavity systems are promising \cite{Lecocq15}, we focus on an implementation of our Bell test with a photonic crystal nanobeam resonator \cite{Chan11, Chan12, Kuramochi10} that distinguishes itself by a very high mechanical frequency $\Omega_m / 2\pi= 5$ GHz. This frequency together with its optical linewidth $\kappa/2\pi < 1$ GHz places this resonator in the resolved sideband regime \cite{Chan12}. The optomechanical coupling rate is large $g_0/2\pi \approx 1$ MHz \cite{Chan12} and mechanical coherence times of the order of 10 to 100 $\mu$s are expected at 4 K and below \cite{Sun13, Chan12}. With a bath temperature $T_{\text{bath}} \approx1.6$K, an initial occupancy of $n_0=0.01$ can be achieved in 100 ns of sideband cooling with 1000 (intracavity) photons corresponding to a peak laser power of 150 $\mu$W \cite{Galland14}. For $T_1=25$ ns and $T_2=50$ ns, we find CHSH$=2.19$ assuming $\eta=90\%$ detection efficiency, fixing $n_-=250$ and optimizing the CHSH value over $n_+=75$,  and over the measurement settings. \\

\paragraph{Perspectives ---} 
Our results show how optomechanical systems can be used to test a Bell inequality. They provide an attractive perspective for the experiment reported in Ref. \cite{Cohen15} where a mechanical system is combined with photon counting techniques. Achieving high overall detection efficiencies is facilitated by photons emitted in a well defined spatial mode which may be coupled into a single mode fiber with a very high-efficiency. Moreover, the wavelength of photons at 1550 nm is an appealing asset to close the locality loophole \cite{Brunner14}. The tremendous progress in optomechanics over the last decade naturally raises the question on whether our proposal might lead to the first loophole-free Bell test. What we can say is that 
our results can also find interesting perspectives in quantum memory experiments \cite{Bussieres13} to certify their proper functioning device-independently. In the context of light storage, off-resonant Raman scattering can be used to create photon-spin wave pairs in atomic ensembles \cite{Duan01}. The spin-wave state can then be mapped to photons using a resonant Raman process -- this mapping is made very efficient thanks to a collective emission. Since the resulting photon-photon state is analog to the photon-photon state described in this letter, the Bell test here used would allow one to certify that this memory operates in the quantum regime without assumptions on the state structure or on the functioning of the measurement devices.\\

\paragraph{Acknowledgements ---} We thank K. Hammerer and S. Hofer for having pointed out the interest of an optomechanical Bell test at a early stage of this work. This work was supported by the Swiss National Science Foundation (SNSF), through the Grant number PP00P2-150579, the Swiss NCCR QSIT, and a SNSF Ambizione Fellowship as well as by the Swiss State Secretariat for Education and Research through the COST Action MP1006.\\

\paragraph{Appendix ---} In the general case where the mechanical oscillator is not in its ground state but in a thermal state with $n_0$ excitations, the joint probability given in Eq. \eqref{joint_proba}  becomes
\begin{eqnarray}
\nonumber
&&P(+1+1|\alpha_1\alpha_2)=
\frac{(1-p)}{\eta' n_0-(\eta'-1) p (\eta+\eta n_0-1)+1} \\
\nonumber
&&\times e^{-\frac{\eta |\alpha_1|^2 (\eta' n_0+(\eta'-1) p+1)+\eta' T \left|\alpha_2\right|^2 (p (\eta+\eta n_0-1)+1)}{\eta' n_0-(\eta'-1) p (\eta+\eta n_0-1)+1}} \\
\nonumber
&&\times e^{-\frac{\sqrt{p}\eta \eta' (n_0+1)\sqrt{T}(\alpha_1^* \alpha_2^*  +\alpha_1 \alpha_2 )}{\eta' n_0-(\eta'-1) p (\eta+\eta n n_0-1)+1}}.
\end{eqnarray}
Similarly for the marginals
$$
P(+1|\alpha_1)=(1-p)\frac{ e^{-\frac{\eta (1-p) |\alpha_1|^2}{ (\eta+\eta n_0-1)p+1}}}{p (\eta+\eta n_0-1)+1},
$$
and
$$
P(+1|\alpha_2)=(1-p)\frac{ e^{-\frac{\eta' (1-p) |\alpha_2|^2T}{(\eta' n_0+(\eta'-1) p+1)}}}{\eta' n_0+(\eta'-1) p+1}.
$$

\end{document}